\documentclass[fleqn,usenatbib]{mnras}

\usepackage{newtxtext,newtxmath}
\usepackage{xcolor, soul}

\usepackage[T1]{fontenc}

\DeclareRobustCommand{\VAN}[3]{#2}
\let\VANthebibliography\thebibliography
\def\thebibliography{\DeclareRobustCommand{\VAN}[3]{##3}\VANthebibliography}

\usepackage{graphicx}	
\usepackage{amsmath}	

\DeclareMathOperator{\Loss}{\mathcal{L}}
\newcommand{\bol}[1]{\boldsymbol{#1}}
\DeclareMathOperator{\w}{\bol{w}}

\DeclareMathOperator{\x}{\mathbf{x}}

\DeclareMathOperator\erf{erf}

\DeclareMathOperator{\p}{\mathrm{pred}}

\title[Deep learning halo bias]{Halo assembly bias from a deep learning model of halo formation}
\author[Lucie-Smith, Barreira \& Schmidt]{
Luisa Lucie-Smith$^{1}$\thanks{E-mail: luisals@mpa-garching.mpg.de},
Alexandre Barreira$^{2,3}$
and Fabian Schmidt$^{1}$ \\
$^{1}$Max-Planck-Institut f{\"u}r Astrophysik, Karl-Schwarzschild-Str. 1, 85748 Garching, Germany\\
$^{2}$Excellence Cluster ORIGINS, Boltzmannstra\ss e 2, 85748 Garching, Germany\\
$^{3}$Ludwig-Maximilians-Universit\"at, Schellingstra\ss e 4, 80799 M\"unchen, Germany \\
}

\date{Accepted XXX. Received YYY; in original form ZZZ}

\pubyear{2015}

\begin{document}
\label{firstpage}
\pagerange{\pageref{firstpage}--\pageref{lastpage}}
\maketitle

\begin{abstract}
We build a deep learning framework that connects the local formation process of dark matter halos to the halo bias. We train a convolutional neural network (CNN) to predict the final mass and concentration of dark matter halos from the initial conditions. The CNN is then used as a surrogate model to derive the response of the halos' mass and concentration to long-wavelength perturbations in the initial conditions, and consequently the halo bias parameters following the ``response bias'' definition. The CNN correctly predicts how the local properties of dark matter halos respond to changes in the large-scale environment, despite no explicit knowledge of halo bias being provided during training. We show that the CNN recovers the known trends for the linear and second-order density bias parameters $b_1$ and $b_2$, as well as for the local primordial non-Gaussianity linear bias parameter $b_\phi$. The expected secondary assembly bias dependence on halo concentration is also recovered by the CNN: at fixed mass, halo concentration has only a mild impact on $b_1$, but a strong impact on $b_\phi$. 
Our framework opens a new window for discovering which physical aspects of the halo's Lagrangian patch determine assembly bias, which in turn can inform physical models of halo formation and bias.
\end{abstract}

\begin{keywords}
large-scale structure of Universe -- dark matter -- galaxies: haloes -- methods: statistical
\end{keywords}



\section{Introduction}
Studying the large-scale structure of the Universe provides valuable information to test our current standard model of cosmology. Since most of the mass in the Universe is in the form of invisible dark matter, observations of galaxies must be used as a proxy to trace the dark matter on cosmological scales. However, galaxies form and reside inside gravitationally bound dark matter halos, and as a result, do not perfectly trace the underlying mass distribution. Thus, it is important to understand the relationship between the large-scale distribution of dark matter and that of dark matter halos in order to make unbiased cosmological inferences using galaxy data.
 
The quantity relating the distribution of halos to the underlying matter distribution is known as \textit{halo bias} (see \cite{Desjacques2018} for a review).
The simplest, most widely used approaches study halo bias only as a function of halo mass and redshift. However, it is well known that dark matter halos of a given mass exhibit bias trends with other secondary properties such as concentration, formation time, spin and shape \citep{Sheth:2004, gao:2005, Wechsler2006, Reid:2010, Mao2018, 2023JCAP...01..023L}. The dependence of the bias on these secondary variables is typically known as \textit{assembly bias}, as they are thought to be related to the assembly history of halos. A robust physical explanation for the existence of secondary bias and its relation to the large-scale environment remains a long-standing question.

The bias of halos is determined by the relative abundance of halos in different large-scale environments. 
A physical picture of halo bias is provided by the ``peak-background split'' (PBS) formalism  \citep{Bardeen1986, ColeKaiser1989, MoWhite1996, ShethTormen1999, ShethMoTormen2001}. The ansatz is that halos form at the locations of rare, high-density peaks of the matter density field at early times. If one decomposes the density field into a large-scale, long-wavelength signal and a short-wavelength signal of the scale at which halos form, the long-wavelength mode acts as a ``background'' to the small-scale fluctuations. For example, in this picture, the number of high-density peaks (or final halos) will be enhanced if these live in an overdense region compared to the average. As a result, a larger number of halos form in overdense regions compared to the mean. Analytic approximations of the PBS formalism based on excursion set and peak theory have clear physical interpretations, but fail to reproduce in detail the bias of halos found in numerical simulations \citep{Seljak2004, Manera2010}.

The PBS approach is a special case of the general ``response bias'' definition, which yields the {\it exact} bias parameters for dark matter halos (see Sec.~3 in \citealt{Desjacques2018}). 
The response bias is defined as the derivative of the number density of halos, or halo mass function, with respect to long-wavelength perturbations, regardless of how this mass function is predicted. That is, unlike PBS, the halo mass function need not be universal or tied to an analytic gravitational collapse model. 
Moreover, the response bias formalism extends beyond perturbations to the background density; it applies to long-wavelength perturbations of any field that can influence structure formation (density, tidal, gravitational potential in case of primordial non-Gaussianity, etc.), each yielding different responses in the final halo abundance. 
The {\it separate-universe} simulation approach provides a way to numerically calibrate the response bias parameters: the bias parameters are obtained as the response of the halo abundance to changes in cosmological parameters that mimic different large-scale backgrounds \citep{Lazeyras2016, baldauf/etal:2015, 2016PhRvD..93f3507L, 2017MNRAS.468.2984P, 2019MNRAS.488.2079B, 2020JCAP...02..005B, 2021MNRAS.503.1473S, 2022JCAP...01..033B}.

However, despite accurate predictions from simulations, it remains unclear which properties of the Lagrangian {\it protohalo} patch that is destined to become a virialized halo determine the halo bias and assembly bias trends. Our goal here is to take steps to bridge the gap between approximate analytic methods that fail to reproduce halo bias quantitatively, and exact numerical techniques from which it is difficult to explain the origin of the bias parameters.

To this end, we build a deep learning framework that connects the properties of the initial peaks that are relevant to the halo formation process with the final large-scale halo bias. This is done by training a deep learning model to learn the mapping between the initial density field and the final properties of dark matter halos; the model is then used to derive the large-scale halo bias as a function of mass, and halo assembly bias as a function of mass and concentration. We focus on concentration since it is a widely explored property of halos, but this approach could be similarly applied to other secondary properties. 
More specifically, the deep learning framework returns predictions for the mass and the concentration of a given population of halos from the initial conditions. The inputs are given by the density field in a sub-region of the initial conditions around the centre of the Lagrangian region of each halo. The trained model is then used to derive 
a number of different halo bias parameters using the response bias definition: by applying large-scale perturbations to the inputs, the CNN predicts the new mass and concentration of the final halos, from which we can measure the bias parameters as the response of the abundance of halos to the large-scale perturbation. We focus on the linear and quadratic density bias parameters $b_1$ and $b_2$, and the linear local primordial non-Gaussianity (PNG) bias parameter $b_\phi$. 

The use of deep learning to create the mapping between the initial conditions and final halo properties allows us to overcome limitations of existing analytic approximations. Machine learning has been recently applied to a variety of problems in structure formation, including predicting halo properties from the initial conditions \citep{LucieSmith2018, LucieSmith2019, LucieSmith2022}, constructing mock dark matter halo catalogues \citep{Berger2019, Bernardini2020}, learning the mapping between the Zel’dovich-displaced and non-linear density fields  \citep{He2019, Jamieson2022} or that between the non-linear density field and the halo distribution \citep{Charnock2020, Ramanah2019}.
More recently, in the context of halo bias, other works have focused on training neural networks to predict the halo bias parameters from initial conditions operators related to the density and tidal shear fields \citep{Wu2022} or from observable properties of simulated galaxies \citep{Sullivan2023}. Our work differs from these in the overall aim and methodology: rather than utilizing machine learning to predict the halo bias parameters directly, we instead derive these via the response bias formalism using a deep learning model that was trained to learn the halo formation process. This allows us to study the connection between the Lagrangian properties that are responsible for halo formation and the bias.

The paper is organized as follows. We first introduce the halo bias parameters studied in this work in Sec.~\ref{sec:biasparams}. In Sec.~\ref{sec:overview}, we provide an overview of our deep learning framework, which consists of a convolutional neural network (CNN) that predicts final halo properties (mass and concentration) from the initial conditions. This is then used as a surrogate model to derive halo bias following the response approach. We describe our $N$-body simulations in Sec.~\ref{sec:sims}, and the construction of the deep learning models that map the initial conditions to halo properties in Sec.~\ref{sec:sims}. We present the halo mass and concentration predictions from the deep learning models in Sec.~\ref{sec:halopreds}. We then move on to deriving the halo bias parameters using the surrogate CNN models; we present the results for the bias parameters as a function of mass in Sec.~\ref{sec:halobias}, and as a function of mass and concentration in Sec.~\ref{sec:assemblybias}. We conclude and discuss future applications of our framework for understanding the origin of assembly bias in Sec.~\ref{sec:conclusions}.

\section{The halo bias parameters}
\label{sec:biasparams}
In this paper, we study the response of halo formation to large-scale perturbations of (i) the matter density and (ii) the primordial gravitational potential with local primordial non-Gaussianity (PNG). 

The simplest, most studied bias parameters are those that connect the density contrast of halos with that of matter at fixed redshift:
\begin{equation}
\delta_h (\mathbf{x}, z)  \equiv \frac{n_h(\mathbf{x}, z)}{\bar{n}_h(z)} - 1 = \sum_n \frac{b_n(z)}{n!} \delta_m^n(\mathbf{x}, z) + \mbox{(non-LIMD terms)}
\end{equation}
where $n_h(\mathbf{x}, z)$ and $\bar{n}_h(z)$ are the local halo number density and its mean, respectively, and we have indicated that this relation ignores all contributions that are not local-in-matter-density (LIMD), such as the tidal field or the gravitational potential. The fractional change in the halo number density is expanded in powers of the matter density contrast $\delta_m$ with coefficients given by the bias parameters $b_n(z)$. In this work, we focus on the present-day bias $b_n \equiv b_n(z=0)$, and consider linear ($n=1$) and quadratic ($n=2$) bias parameters. 

The other bias parameter we consider, $b_\phi$, enters the halo number density contrast as $\delta_h(\mathbf{x}, z) \supset b_\phi(z) f_\mathrm{NL}\phi(\mathbf{x})$ \citep{mcdonald:2008, assassi/baumann/schmidt}, where $\phi$ is the primordial gravitational Bardeen potential and $f_\mathrm{NL}$ is a dimensionless parameter defined as \citep{Komatsu2001}
\begin{equation}
\phi(x) = \phi_G(x)  + f_\mathrm{NL} \left[ \phi_G(x) ^2 -  \langle \phi_G(x) ^2\rangle \right],
\end{equation}
and $\phi_G(x)$ is a Gaussian random field. This way to parametrize departures from perfect Gaussianity of the distribution of $\phi$ is called {\it local primordial non-Gaussianity} (PNG). Canonical single-field slow-roll inflation predicts $\phi$ to be essentially Gaussian \citep{Bardeen1986}. A measurement of $f_\mathrm{NL}$ would thus constitute direct evidence for non-standard inflationary physics in the early Universe \citep{maldacena:2003, Bartolo2004, 2004JCAP...10..006C, Tanaka:2011aj, conformalfermi}. The current best constraints come from measurements of the bispectrum or three-point correlation function of the cosmic microwave background (CMB) by the Planck satellite, which set $f_\mathrm{NL} = -0.9 \pm 5.1\ (1\sigma)$ \citep{Planck2018fNL}. The large-scale galaxy power spectrum is also a very powerful probe of local PNG through a distinctive scale-dependent signature that scales as $\propto b_{\phi} f_\mathrm{NL}/k^2$ \citep{Dalal2008, Slosar2008}. The perfect degeneracy between $b_\phi$ and $f_\mathrm{NL}$ on this effect shows that a good knowledge of $b_\phi$ is important for robust constraints on $f_\mathrm{NL}$ \citep{2022JCAP...11..013B}.

Physically, local PNG generates a squeezed-limit bispectrum, which induces a modulation of the small-scale power spectrum $P_{\phi\phi}(k)$ of $\phi$ by long-wavelength modes $\phi_{\rm L}$ \citep{Slosar2008, Matarrese2008}. Concretely, we can write \citep{Desjacques2018}
\begin{equation}
P_{\phi \phi} (k, z | \mathbf{x}) = P_{\phi \phi} (k, z) \left[ 1 + 4 f_\mathrm{NL} \phi_{\rm L}(\mathbf{x})\right],
\end{equation}
where the subscript ${}_{\rm L}$ emphasises this is a perturbation with a wavelength much larger than $1/k$. This equation shows that the effect of local PNG is equivalent to a change in the amplitude of the primordial power spectrum $A_s \rightarrow A_s \left[ 1 + \delta A_s \right]$ where $\delta A_s = 4 f_\mathrm{NL} \phi_{\rm L}(\mathbf{x})$. The parameter $b_\phi$ is thus defined as the response of the halo number density to a large-scale perturbation $f_\mathrm{NL}\phi_{\rm L}$, or equivalently by the separate-universe argument, a change in $A_s$, i.e., 
\begin{equation}
b_\phi \equiv \frac{\partial \ln  n_h}{ \partial (f_\mathrm{NL} \phi_{\rm L}(\mathbf{x}))} = 4 \frac{\partial \ln  n_h}{ \partial \delta A_s}.
\end{equation}

\begin{figure*}
     \centering
         \includegraphics[width=\textwidth]{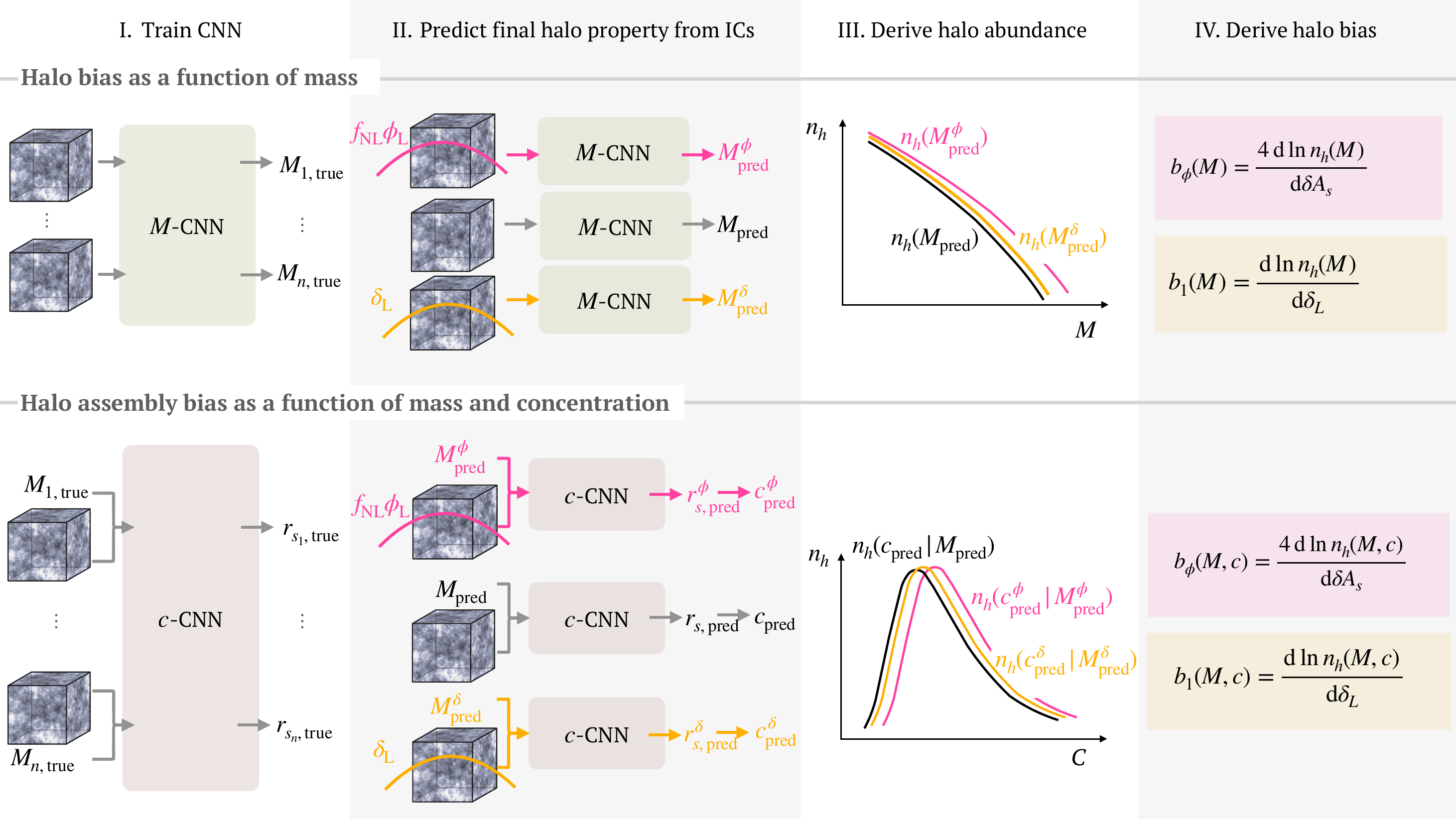}
            \caption{An overview of the employed pipeline to derive halo bias from a surrogate CNN model that predicts individual halo properties from the halos' initial Lagrangian patch. Two CNN are trained to predict mass and concentration (top/bottom panel), respectively, of individual halos from the initial density field around the protohalo centre (step I). We mimic a change in the background density field by adding a constant density perturbation $\delta_{\rm L}$ to the input density field; the trained CNN then predict the modified mass and concentration associated with that halo. We also mimic a change in the primordial gravitational potential by changing the variance of the input density field by $\delta A_s  = 4 f_\mathrm{NL}\phi_{\rm L}$; the trained CNN again predict new mass/concentration values. We measure the abundance of halos as a function of mass (top) and concentration at fixed halo mass (bottom; step III), and their responses to the large-scale perturbations, thus yielding the halo bias parameters (step IV).}
    \label{fig:illustration}
\end{figure*}

\section{Overview of the deep learning framework}
\label{sec:overview}

Our goal is to develop a deep learning framework that can be used to study the physical connection between the initial density peaks from which halos form and halo assembly bias.
The framework consists of a deep learning model that predicts final halo properties from the initial conditions; the CNN is then used as a surrogate model to derive the halo bias. We first focus on deriving halo bias as function of mass, and then extend the framework to halo assembly bias as a function of mass and concentration. An overview of our pipeline is illustrated in Fig.~\ref{fig:illustration}. We start with a description of the steps involved in deriving halo bias a function of mass (top row, Fig.~\ref{fig:illustration}), and halo assembly bias as a function of mass and concentration (bottom row, Fig.~\ref{fig:illustration}).

\subsection*{Step I: Training the CNN model}
The first step consists of training one CNN to predict halo mass ($M$-CNN model), and a different CNN to predict the Navarro-Frenk-White (NFW; \citealt{Navarro1997}) scale radius $r_s$ ($c$-CNN model). The former will be used to derive the halo bias a function of mass, the latter the halo assembly bias as a function of mass and concentration. The inputs are given by the initial density field in a large sub-region of the simulation around the centre-of-mass of each protohalo patch; additionally, the $c$-CNN also takes as input the halo mass (see Sec.~\ref{sec:inputs} for more details and motivation for this choice). The networks have no explicit knowledge of halo bias during the training process; the $M$-CNN and $c$-CNN learn to identify the aspects of the initial density field which impact only the final halo mass and concentration, respectively.

\subsection*{Step IIa: Predicting halo mass and concentration}
The second step involves making predictions for the final halo properties for halos of an independent simulation, using the trained CNN models. We use the $M$-CNN model to predict the halo mass ($M_{\p}$), and the $c$-CNN model to predict the halo scale radius. In the prediction phase, the $c$-CNN model takes as input the predicted mass from $M$-CNN, $M_{\p}$; the scale radius predictions are turned into concentration predictions using $c=r_{\rm 200m}/r_s$, where $r_{\rm 200m} = (3M_{\p}/4 \pi \rho_{\rm 200m})^{1/3}$ and $M_{\p} \equiv M_{200m}$ is defined as the mass enclosed within a sphere containing a density ($\rho_{\rm 200m}$) that is 200 times the mean matter density at $z=0$.

\subsection*{Step IIb: Adding long-wavelength perturbations to the inputs}
The halo bias parameters describe the response of the halo number density to a set of long-wavelength perturbations; we therefore inject long-wavelength perturbations to the input density field of every halo and predict the resulting mass and concentration of the halo using the trained CNN models.
To mimic a change in the background density field, we add a constant density perturbation $\delta_{\rm L}$ to the density field in every input sub-box; given these perturbed inputs, the trained $M$-CNN predicts new halo masses $M_{\p}^{\delta}$, and the trained $c-$CNN predicts new concentration values $c_{\p}^{\delta}$. To mimic a perturbation to the primordial potential in local PNG cosmologies, we modify the variance of the field by multiplying the density in every input sub-box by $\sqrt{1 + \delta A_s}$; given these perturbed inputs, the trained $M$-CNN and $c$-CNN models again yield new mass and concentration predictions for the halos, $M_{\p}^\phi$ and $c_{\p}^\phi$.

\subsection*{Step III: Inferring halo abundance}
The third step consists in turning the predictions of individual halos into the summary statistic of interest i.e., the number density of halos as function of the predicted property. We histogram the values $M_{\p}$ of individual halos in predefined mass bins to obtain the halo abundance as a function of mass, $n_h(M)$; similarly, we histogram the $c_{\p}$ values of halos of a given predicted mass in predefined concentration bins to obtain the halo abundance as a function of concentration conditioned on mass, $n_h(c|M)$. We compute the halo abundances for all three sets of predicted properties: those from inputs with no perturbations ($M_{\p}$, $c_{\p}$), those from inputs with an added background density ($M_{\p}^{\delta}$, $c_{\p}^{\delta}$), and those from inputs embedded in a potential perturbation ($M_{\p}^\phi$, $c_{\p}^\phi$).

\subsection*{Step IV: Deriving halo bias}
The final step involves deriving the halo bias parameters by measuring the response of the number density of halos to the density $\delta_{\rm L}$ and primordial potential perturbations $\phi_{\rm L}$ as
\begin{equation}
b_1(X) = \frac{\mathrm{d} \ln n_h(X)}{\mathrm{d} \delta_{\rm L}}
\label{eq:b1deriv}
\end{equation}
and 
\begin{equation}\label{eq:bphidef}
b_\phi(X) = \frac{4 \, \mathrm{d} \ln n_h(X)}{\mathrm{d} \delta A_s},
\end{equation}
where $X$ is a generic halo property. When computing the halo bias as a function of mass, $X=M_{\p}$; when measuring the halo assembly bias as function of mass and concentration, $X=c_{\p}|M_{\p}$. In practice, we use a generalized form of Eqs.~\eqref{eq:b1deriv} and \eqref{eq:bphidef} to higher order in $\delta_{\rm L}$ and $\delta A_s$, which we describe in more detail in Sec.~\ref{sec:halobias}.

\section{Simulations}
\label{sec:sims}
We generated the training and testing data from five gravity-only $N$-body simulations produced with \texttt{AREPO} \citep{Arepo2020}, consisting of a box of size $\mathrm{L}= \, 560 \, \mathrm{Mpc} \, h^{-1}$ and $N = 1250^3$ matter tracer particles evolved from $z=127$ to $z=0$. Four simulations were used for training and one for validation and testing. We made use of pynbody \citep{pynbody} and \texttt{nbodykit} \citep{Hand2018} to analyse the information contained in the simulation snapshots. Each simulation is based on a different realization of a Gaussian random field drawn from the initial power spectrum of density fluctuations, generated using \texttt{N-GenIC} \citep{2015ascl.soft02003S}.

Dark matter halos were identified at $z=0$ using the \texttt{ROCKSTAR} halo finder \citep{Behroozi2013}, a phase-space halo-finder that uses an adaptive hierarchical refinement of six-dimensional friends-of-friends (FoF) and one time dimension to track merged structure over time. We used the following halo properties provided by \texttt{ROCKSTAR}: the halos' mass and scale radius. We focused on high-mass halos within the mass range $M_{\rm 200m} \in [10^{13}, \, 5\times10^{14}] \, M_{\odot} \, h^{-1}$. The minimum mass scale of $10^{13} \,M_{\odot} \, h^{-1}$ was chosen in order to fully resolve the inner profile of the halos and obtain reliable concentration estimates; the choice of a maximum mass scale of $5\times10^{14} \,M_{\odot} \, h^{-1}$ ensured the training set would not be too severely affected by small-number statistics at the high-mass end. The scale radius was measured from the ratio between the maximum of the circular velocity $V_\mathrm{max}$ and $V_\mathrm{200m}$, the circular velocity at $r_\mathrm{200m}$, assuming an NFW profile  \citep{Prada2012, Klypin2011}. 
Given the scale radius, we infer the concentration $c = r_\mathrm{200m}/r_s$. We also made use of \texttt{ROCKSTAR} to identify the bound particles assigned to each halo by the halo finder.

We also considered a set of separate-universe simulations to measure the bias parameter $b_\phi$, which we compare to the CNN-derived $b_\phi$. Concretely, for the same initial conditions realization as the simulation we use for testing the CNN, we ran two additional simulations with the same cosmological parameters except for $A_s \to A_s\left(1 + \delta_{A_s}\right)$, with $\delta_{A_s} = \pm 0.05$. These simulations were used to evaluate Eq.~(\ref{eq:bphidef}) using finite differences, and obtain the ``true'' $b_\phi$ to compare the CNN predictions with. Note that in the limit of weak PNG ($f_{\rm NL} \lesssim 10$), it is sufficient to compute $b_\phi$ from simulations with Gaussian initial conditions, as the halo mass function is only weakly affected by this amount of primordial non-Gaussianity. We obtain the true values of $b_1$ as $b_1(z) = {\rm lim}_{k\to0} P_{hm}(k,z)/P_{mm}(k,z)$, where $P_{mm}$ is the matter power spectrum of the test simulation and $P_{hm}$ the halo-matter cross-power spectrum of halos in some mass and concentration bin. We emphasize these true bias values are used {\it only} for testing and {\it never} to train the CNN.

\section{Building CNN for the initial conditions-to-halo properties mapping}

\subsection{The inputs and outputs of the $M$-CNN and $c$-CNN models}
\label{sec:inputs}
We consider two CNN models that predict two halo properties at $z=0$ from the initial conditions, respectively: one that predicts the halo mass $M_\mathrm{200m}$, which we denote $M-$CNN, and one that predicts the halo scale radius, from which we then infer the halo concentration $c$, which we denote $c-$CNN. More specifically, the output of the $M-$CNN is given by $\log_{10}(M_\mathrm{200m} / [\mathrm{M_\odot}/h])$; the output of the $c-$CNN model is given by $\log_{10}(r_s/[ \mathrm{kpc}/h])$. For both models, the outputs are rescaled to the range $ \left[-1, 1 \right]$.

The input to both the $M-$CNN and $c-$CNN models for any given halo is given by the initial density field $\delta(\mathbf{x})$ in a cubic sub-region of the initial conditions of the simulation, centred on the centre-of-mass of the halo's Lagrangian patch. The centre-of-mass of the protohalo was identified by tracking the particles that make up the $z=0$ halo back to their initial positions, and computing their centre-of-mass. The input sub-region covers a ($34 \, \mathrm{Mpc} \, h^{-1})^3$  sub-volume of resolution $N = 77^3$. The volume of the input sub-region was chosen to be several times the halos' Lagrangian radius $R_L \sim 3 - 10 \, \mathrm{Mpc}/h$, while the resolution was chosen so that the grid scale would be a small fraction of $R_L$. We varied both the volume and the resolution of the inputs to find the values which yielded the best-performing model. The density field within each input sub-region was estimated from the particle positions; specifically, we estimated the density at the location of each particle following a smoothed-particle hydrodynamics (SPH) procedure where the SPH kernel smoothing length depends on each particle's 32 nearest neighbours, and then projected this on to the regular grid of the input sub-box. Finally, the density is turned into the density contrast $\delta ((\mathbf{x}) = (\rho(\mathbf{x}) - \bar{\rho}_m)/ \bar{\rho}_m$, where $\bar{\rho}_m$ is the mean matter density of the simulation.

The $c-$CNN model additionally takes as input the mass of the halo. The latter is given by the ground-truth halo mass during training, but replaced by the predicted halo mass from the $M-$CNN model during testing. If our interest were solely to predict concentration (or scale radius) from the initial conditions, the additional mass input to $c-$CNN model would not be needed; the $c-$CNN model is capable of predicting halo concentrations without any prior knowledge about halo mass. However, since we plan to use the CNN predictions to measure the response of the halo concentration to a large-scale perturbation in the initial conditions, we must first isolate the change in halo mass to that same large-scale perturbation. In other words, we must isolate the change in the halo profile to the change in mass due to the same large-scale perturbation in the inputs. Adding halo mass as input to the $c-$CNN model implies that the response of the halo mass to the added large-scale perturbation is incorporated by such input, so that any residual response in the predicted concentration is due to a change in the halo profile alone. Other design choices that isolate the response of the mass to that of the profile exist, as for example predicting halo mass and concentration simultaneously with a single CNN. We found that such a CNN architecture was more difficult to train compared to the two-network design that we adopt in this work.

\subsection{The CNN models}
Having specified the training data, we next turn to defining the deep learning model that we use. The $M$-CNN and $c$-CNN models have very similar architectures. The  models consist of five convolutional layers and four fully-connected layers. The convolutional layers adopt three-dimensional kernels, so that they can be applied to the input 3D initial density field; all kernels have size $3 \times 3 \times 3$. The convolutions were performed with 32, 32, 64, 64, 64 kernels with stride$\,=1$ for the five convolutional layers, respectively. All convolutional layers (but the first one) are followed by max-pooling layers; their output is then used as input to the non-linear leaky rectified linear unit (LeakyReLU) \citep{Nair2010} activation function. As more convolutional layers are stacked on top of each other, the CNN learns more global, large-scale features from the input data.

After the last convolutional layer, the output is flattened and passed on to a series of four fully-connected layers, each made of 256, 128, 128 and 1 neuron, respectively. The $c$-CNN model also takes mass as an additional input; this is passed on to the first fully-connected layer together with the flattened output of the last convolutional layer. The non-linear activation function of the first three fully-connected layers is the same LeakyReLU activation as that used in the convolutional layers, whereas the last layer has a linear activation in order for the output to represent the halo mass or the scale radius.

\subsubsection{The loss function}
Training the CNN requires optimizing the parameters of the model, $\w$, that minimize a loss function measuring how closely the predictions are to their respective ground truths for the training data. The goal of the neural network is to maximize the posterior distribution $p \left( \w \mid  \mathcal{D} \right) = p \left( \mathcal{D} \mid  \w \right) p \left( \w \right)$, where $p \left( \mathcal{D} \mid  \w \right)$ is the likelihood of the training data $\mathcal{D}$ given the model weigths $\w$ and $p \left( \w \right)$ is the prior over the weights. The loss function, $\Loss$, is then given by
\begin{equation}
\Loss = - \ln \left[ p(\w \mid \mathcal{D}) \right] = - \ln \left[ p \left( \mathcal{D} \mid  \w \right) \right]  - \ln \left[ p(\w) \right],
\label{eq:posterior_weights}
\end{equation}
where the first is the likelihood term, or predictive term $\Loss_{\p}$, and the second is the prior term, or regularization term $\Loss_\mathrm{reg}$. 

The training data $\mathcal{D}$ is comprised of pairs of inputs $\tilde{\x}$ and ground-truth values $d = d \left( \tilde{\x} \right)$ for every halo in the training set.
The likelihood function describes the distribution of ground truths $d$ (i.e. the actual mass or scale radius of the halos) for a given value of predicted output, which we denote as $y = y \left( \tilde{\x}, \w \right)$. Since we restrict the ground truths to the values $\left[-1, 1\right]$, we have the additional constraint of a top-hat selection function $S$ over the ground truth variable, $p \left( S \mid d \right) = \Theta \left(1 - d \right) \Theta \left( d +1 \right)$, where $\Theta$ is the Heaviside step function. The loss function for any likelihood distribution under selection by a top hat selection function takes the form \citep{Lucie-Smith2020}
\begin{align}
\begin{split}
\Loss_{\p} = & - \ln \left[ p \left(d \mid y, S \right) \right] \\
= &- \ln \left[ p \left(d \mid y \right) \right] + \ln \left[  p \left(d \leq 1 \mid y \right) - p \left(d \geq -1 \mid y \right) \right],
 \label{eq:likselection}
 \end{split}
\end{align}
where the first term is the typical log-likelihood term without any selection, and the second term comes from the selection function constraint.

We assume a Gaussian likelihood, meaning that the loss takes the specific form
\begin{equation}
\Loss_{\p} = \frac{1}{N} \sum_{i=1}^N \frac{1}{2} \left( \frac{d_i - y_i}{\sigma} \right)^2 + \ln \left[ \erf{} \left( \frac{1 - y_i}{\sqrt{2} \sigma}\right) - \erf \left( \frac{-1 - y_i}{\sqrt{2} \sigma} \right) \right],
\label{eq:gauss_sel}
\end{equation}
where $N$ is the number of halos in the training set, and the standard deviation $\sigma$ is a free parameter that must be set prior to training. We choose $\sigma=0.2$, but have verified that the training is insensitive to the specific choice of $\sigma$. Note that the loss in Eq.~\eqref{eq:likselection} implicitly assumes that both $d, y \in [-1,1]$. Although this is true for $d$, it is not strictly true for $y$ since the predicted values of the CNN can in principle assume any value. This was corrected by introducing an additional super exponential term to $\Loss_{\p}$ for predicted values $|y| > 1$, thus strongly disfavouring predictions in that regime.

The regularization term in Eq.~\eqref{eq:posterior_weights}, $\Loss_\mathrm{reg}$, is intended to simultaneously (i) improve the optimization during training by preventing the algorithm from overfitting the training data and (ii) compress the neural network model into the smallest number of parameters without loss in performance. We adopt Gaussian priors for the weights of the convolutional layers which yields L2 regularization, and Laplacian priors for the weights of the dense layers which yields L1 regularization. These choices promote small values for the weights thus reducing the model's ability to overfit, while the choice of L1 regularization has the additional benefit of favouring sparsity in the neurons of the fully-connected layers.

\begin{figure*}
     \centering
         \includegraphics[width=0.49\textwidth]{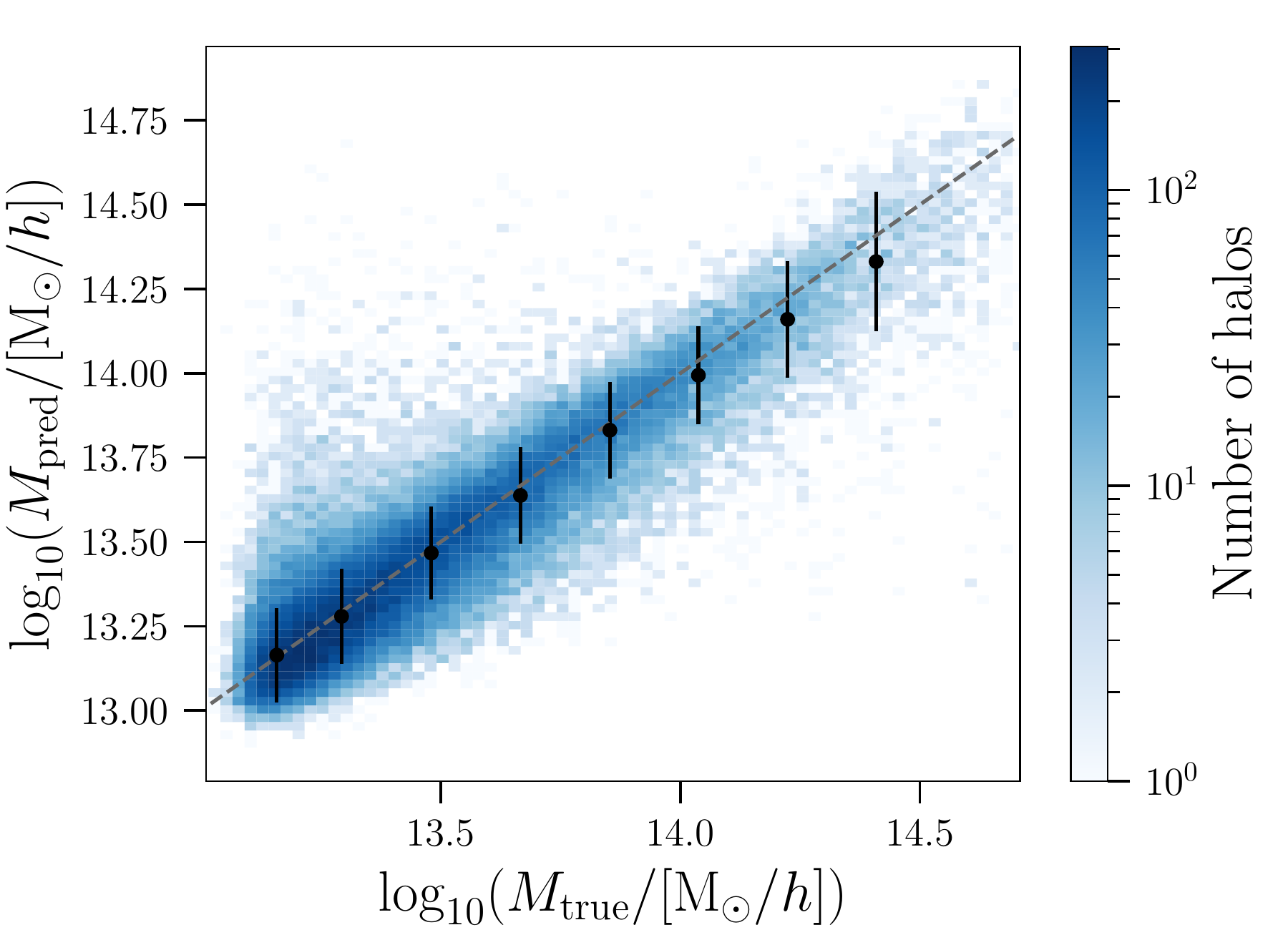}
         \includegraphics[width=0.49\textwidth]{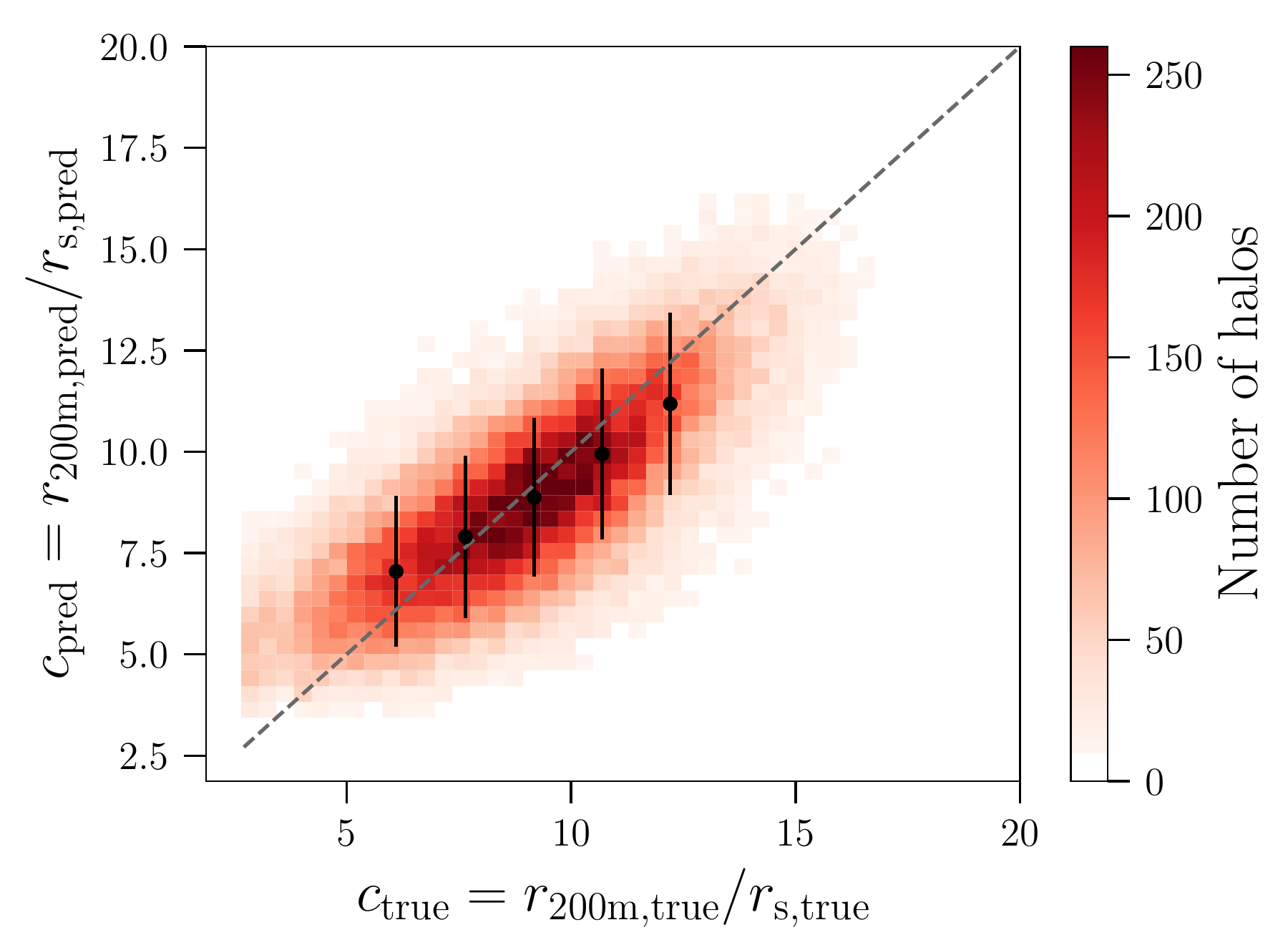}
            \caption{\textit{Left}: A CNN (which we denote $M$-CNN) is trained to predict the final mass of a halo given the initial density field around the centre-of-mass of the protohalo region. The results are shown as two-dimensional histograms in predicted-vs-true space; the errorbars show the median and standard deviation in bins of true mass. \textit{Right}: A different CNN (which we denote $c$-CNN) is trained to predict the final concentration of a halo given as inputs (i) the initial density field around the centre-of-mass of the protohalo region and (ii) the final halo mass. For the latter, we use the ground-truth halo mass during training, and that predicted by the $M$-CNN model during testing. The results are also shown as two-dimensional histograms in predicted-vs-true space. The variance in the concentration predictions is larger, as predicting concentration from the initial conditions is a more difficult task than predicting halo mass.}
    \label{fig:CNNpred}
\end{figure*}

\section{CNN predictions}
\label{sec:halopreds}
We applied the trained CNN models to halos from an independent simulation not used for training. We show the results in Fig.~\ref{fig:CNNpred}, where we compare the predictions made by the CNN to the ground truth values of the halo properties (mass and concentration). The predictions are shown as two-dimensional histograms in predicted-ground truth space; the left panel shows the results for mass and the right panel for concentration. Note that the colorbar in the left (right) panel reflects log-spaced (linearly spaced) counts of halos in each histogram bin. The errorbars show the median and standard deviation in bins of ground-truth values. The grey dashed line in both panels shows $y=x$ and represents the idealized case of 100\% accuracy. 

We find good predictions for the halo mass, meaning that the CNN has learnt to identify features in the initial conditions which contain relevant information about the final halo mass. The accuracy, in terms of the average standard deviation of the residuals in different mass bins, is $\sim 1\%$. In \cite{Lucie-Smith2020}, a CNN was also trained to predict final halo masses from the initial conditions. In that work, the mapping was done for every \textit{particle}, so that the input was given by the density field around each particle's initial position and the output by the mass of the halo to which that particle belongs at $z=0$; in this work, the mapping is done for every \textit{halo}, so that the input is given by the density field around the protohalo's centre-of-mass. The additional information provided to the CNN about the centre-of-mass of the protohalo yields significantly improved mass predictions compared to those in \cite{Lucie-Smith2020}.

The concentration predictions have a larger variance than those for mass (accuracy $\sim 20\%$), reflecting the fact that predicting concentration from the initial conditions is a notoriously difficult task compared to predicting halo mass. This is because the final concentration of a halo is strongly affected by the assembly history of the halo, which is not readily available in the form of features of the initial conditions alone. 
Note that the CNN is not provided any direct information about the evolution of the density field over time.
Moreover, the accuracy in the concentration predictions is affected by both the accuracy of the $M_\mathrm{200m}$ predictions, which enters via $r_\mathrm{200m}$, and the accuracy of the $r_s$ predictions. If we remove the uncertainty in the mass by defining the predicted concentration values as $c_\mathrm{pred} = r_\mathrm{200m, true}/r_\mathrm{s, pred}$, the residuals shrink by about $20 - 30 \%$. 

The uncertainties in the predictions reflect how predictive the features of the initial conditions discovered by the CNN are about the final halo property; this will in turn affect the ability of the CNN to derive the correct halo bias.

\section{Deriving halo bias from the CNN mass predictions}
\label{sec:halobias}

\begin{figure*}
     \centering
         \includegraphics[width=\textwidth]{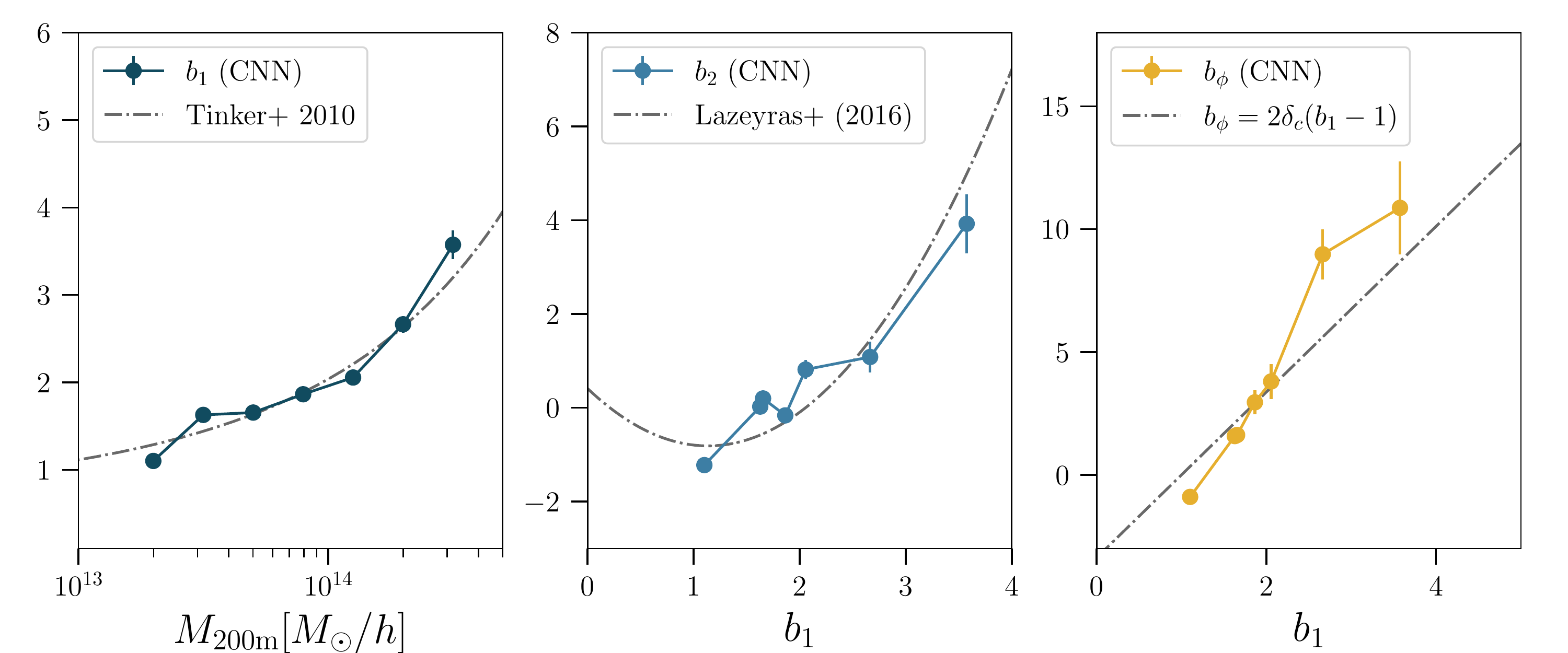}        
            \caption{Linear density bias $b_1$ (\textit{left}), quadratic density bias $b_2$ (\textit{middle}) and local PNG bias $b_\phi$ (\textit{right}) measured using the $M$-CNN predictions for halo mass. In the middle and right panels, the bias values are plotted for the same mass bins shown on the left. For comparison, we show the \citet{Tinker2010} fitting function for $b_1(M)$, the \citet{Lazeyras2016} fitting function for $b_2 (b_1)$, and the universality relation for $b_\phi (b_1)$. The CNN is able to reproduce the expected values of the bias parameters, despite not being specifically trained to do so: the $M$-CNN is trained only on the mapping between initial conditions and final halo mass.}
    \label{fig:b1}
\end{figure*}

Next, we use the trained networks as surrogate models to derive halo bias. As mentioned in Sec.~\ref{sec:overview}, halo bias is defined as the response of the halo abundance to a large-scale perturbation. In this section, we focus on the response of the halo abundance as a function of mass (halo bias), and in the next Sec.~\ref{sec:assemblybias} we discuss the response as a function of mass and concentration, thus probing halo assembly bias. In each case, we consider two types of large-scale perturbations: (i) perturbations to the background matter density which let us study the bias parameters $b_1$ and $b_2$, and (ii) to the primordial gravitational potential, which let us study $b_\phi$. 

\subsection{Linear and second-order density halo bias, $b_1$ and $b_2$}
\label{sec:b1mass}

To mimic a change in the background density field to measure $b_1$ and $b_2$, we add a constant density perturbation $\delta_{\rm L}$ to the input initial density field of every halo in the test set. We use $\delta_{\rm L} = \{ \pm0.5,  \pm0.4,  \pm0.3,  \pm0.2,  \pm0.1,  \pm0.07, \pm0.05, \pm0.02, \pm0.01 \}$, where $\delta_{\rm L}$ is the present-day linearly extrapolated matter density contrast. This is then rescaled by the growth function $\delta_{\rm L}(z=127) = D(z=127)/D(z=0) \, \delta_{\rm L}$ before adding it to each initial conditions input density field. The new inputs are then passed to the CNN, which returns new mass predictions for each halo. We use the predicted masses to evaluate the halo mass function $n_h(M_{\p})$ for all $\delta_{\rm L}$ values, including the $\delta_{\rm L} = 0$ case. We measure $n_h(M_{\p})$ in bins from $\log_{10} M_\mathrm{min}=13.2$ to $\log_{10} M_\mathrm{max}=14.6$  with log-spacing $\Delta \log_{10} M = 0.2$. The change in the number density of halos in response to the added background density in a mass bin centred on $M$ is given by 
\begin{equation}
\delta_h(M) = \frac{n_h(M^{\delta}_{\p})}{n_h(M^{0}_{\p})} -1,
\label{eq:deltah}
\end{equation}
where $n_h(M^{\delta}_{\p})$ and $n_h(M^{0}_{\p})$ are the number densities of halos inferred from the CNN predicted masses given the input density field with and without an added perturbation $\delta_{\rm L}$, respectively.

For every mass bin, we estimate the bias parameters using a polynomial expansion in $\delta_{\rm L}$,
\begin{equation}
\delta_h(M_{\p}) = \sum_{i=1}^N \frac{1}{n!}b_n^L(\delta_{\rm L})^n,
\label{eq:polybias}
\end{equation}
where $b_n^L$ are the nth-order Lagrangian bias parameters. We use a 4th order polynomial in $\delta_{\rm L}$ to fit Eq.~\eqref{eq:polybias} and extract the best-fitting $b_1^L$ and $b_2^L$ parameters. Finally, the Eulerian bias parameters are obtained from the Lagrangian ones via $b_1(M) \equiv b_1^L(M) + 1$ and $b_2(M) \equiv  \frac{8}{21} b_1^L(M) + b_2^L(M)$.  We bootstrap the mass prediction data 40 times and repeat the process to derive $b_1$ and $b_2$ for each bootstrapped realization; this yields the mean and standard deviation of the distribution of $b_1$ and $b_2$ values.

Figure \ref{fig:b1} shows the results for the Eulerian halo bias parameters derived using the CNN framework. The left and middle panels show $b_1(M)$ and $b_2(b_1)$. The results for $b_1(M)$ are in very good agreement with the fitting function of \citet{Tinker2010} shown by the dot-dashed line. The \citet{Tinker2010} fitting function was calibrated to reproduce the $b_1$ values of halos (identified using the spherical overdensity algorithm) in $N$-body simulations; in their work, $b_1$ is measured as the ratio of the halo power spectrum to the linear dark matter power spectrum. The \cite{Tinker2010} fitting function has an accuracy of $5-10\%$. Despite not being explicitly trained to learn about halo bias, our deep learning framework is capable of deriving $b_1$ with the same level of accuracy as the directly calibrated fitting function. The CNN learns solely about the features of the initial density field that are responsible for the final mass of a halo; based on this, it derives the correct $b_1$ by predicting the correct linear response of the halo mass to a change in the initial background density.

In the middle panel of Fig.~\ref{fig:b1} , we compare our results for $b_2(b_1)$ to the fitting formula of \citet{Lazeyras2016} given by $b_2 (b_1) = 0.412 - 2.143 \, b_1 + 0.929 \, b_1^2 + 0.008 \, b_1^3$. The latter was found by fitting to the measured values of $b_2$ and $b_1$ for dark matter halos in $N$-body simulations, obtained using the separate-universe approach. These measurements were also found to be in good agreement with the bias measured using the halo-matter power spectrum and bispectrum \citep{Lazeyras2016}. We find that the CNN is able to also reproduce the quadratic response of the halo number density to a large-scale perturbation in the matter density field, meaning that the features extracted by the CNN are sufficiently predictive to capture non-linear interactions with the large-scale environment.

\subsection{Linear local primordial non-Gaussianity (PNG) bias, $b_\phi$}
To mimic a large-scale gravitational potential perturbation to measure $b_\phi$, we multiply the initial density field within each input sub-box by a factor of $\sqrt{1 + \delta A_s}$, where we use $\delta A_s = \{\pm0.1, \pm0.05, \pm0.01\}$. 
The new inputs are then passed to the CNN, which returns new mass predictions for each halo, and in turn a new inferred halo mass function. Analogously to Eq.~\eqref{eq:deltah}, the change in halo abundance is given by
\begin{equation}
\delta_h(M) = \frac{n_h(M^\phi_{\p})}{n_h(M^0_{\p})} -1,
\label{eq:deltahbphi}
\end{equation}
where $n_h(M^\phi_{\p})$ is the number density of halos inferred from the CNN predicted masses given inputs under a long-wavelength potential perturbation $f_{\rm NL}\phi_{\rm L}$. The first-order $b_\phi$ parameter is inferred by fitting a polynomial expansion in $\delta A_s$, such that
\begin{equation}
\delta_h(M) = \sum_{i=1}^N \frac{1}{n!}b_n^\phi(\delta As)^n,
\label{eq:polybiasbphi}
\end{equation}
where $b_n^\phi$ are the nth-order PNG bias parameters. We use a 4th order polynomial in $\delta A_s$ to fit Eq.~\eqref{eq:polybiasbphi} and extract the best-fit $b_n^\phi$ parameters. The final PNG bias parameter is given by $b_\phi = 4 b_1^\phi$; the factor of 4 arises because $\delta A_s = 4 f_\mathrm{NL} \phi_{\rm L}$.

The right panel of Fig.~\ref{fig:b1} shows the results for $b_\phi(b_1)$ inferred from the CNN predictions. Although $b_\phi$ and $b_1$ describe different physical responses, it is common to parametrize $b_\phi$ in terms of $b_1$ since observational constraints rely on priors over this relation rather than on $b_\phi$ alone. We compare the CNN results with the popular universality relation $b_{\phi} = 2\delta_c(b_1 -1)$, which follows from assuming universality of the halo mass function, where $\delta_c = 1.686$ is the (linearly extrapolated to z = 0) threshold overdensity for spherical collapse. This relation has been found to slightly overpredict the $b_\phi(b_1)$ relation of halos in $N$-body simulations for $b_1 > 2$ \citep{grossi/etal:2009, desjacques/seljak/iliev:2009, 2010MNRAS.402..191P, 2011PhRvD..84h3509H, 2017MNRAS.468.3277B, 2020JCAP...12..013B, 2022JCAP...01..033B}, but it is sufficient as a benchmark against which to compare our CNN results. Concretely, the CNN predictions broadly recover the expected linear relation, except for the lowest/highest $b_1$ values (or equivalently lowest/highest halo mass bins) which slightly underestimate/overestimate the prediction from the universality relation.

In tests of the performance of the CNN, we found that its ability to predict $b_\phi$ is dependent on the spatial resolution of the density field used as input to the $M$-CNN model. We find that we require at least a resolution of $77^3$ in a sub-box of size $L=34 \, \mathrm{Mpc}\,/h$, which corresponds to a spatial resolution of grid size $l \sim 0.44 \, \mathrm{Mpc}\,/h$ (comoving). When the resolution is reduced to $51^3$ for the same sub-box size, i.e. $l \sim 0.67 \, \mathrm{Mpc}\, /h$, $b_\phi$ is systematically underestimated for all values of $b_1$ by $\sim 50 \%$ for all halo masses. If we increase the resolution to $91^3$, the results remain consistent with those using inputs of $77^3$ resolution, meaning that a spatial resolution of $l \sim 0.44 \mathrm{Mpc}\,/h$ is sufficient to capture the relevant information. On the other hand, we find that our results do not change if we increase the size of the input sub-box further. Physically, this implies that, for our mass range, the $b_\phi$ parameter is probing the response to the change of the variance of modes on scales that are at least $l \lesssim 0.7 \, \mathrm{Mpc}\, /h$. In contrast, the results for $b_1$ and $b_2$ were identical for inputs of lower resolution, meaning that the peak properties that determine these two bias parameters are on larger scales $l \gtrsim 0.7 \, \mathrm{Mpc}\, /h$.

Overall, the results in this section demonstrate that the CNN is able to predict the correct response to large-scale perturbations without being explicitly trained to do so; the features extracted by the CNN to determine halo mass respond to the large-scale perturbations in such a way that the correct halo bias is recovered. Therefore, the model makes use of the intricate connection between local peak properties and the large-scale environment to determine the correct final halo bias.

\section{Halo assembly bias}
\label{sec:assemblybias}
\begin{figure*}
     \centering
         \includegraphics[width=0.95\textwidth]{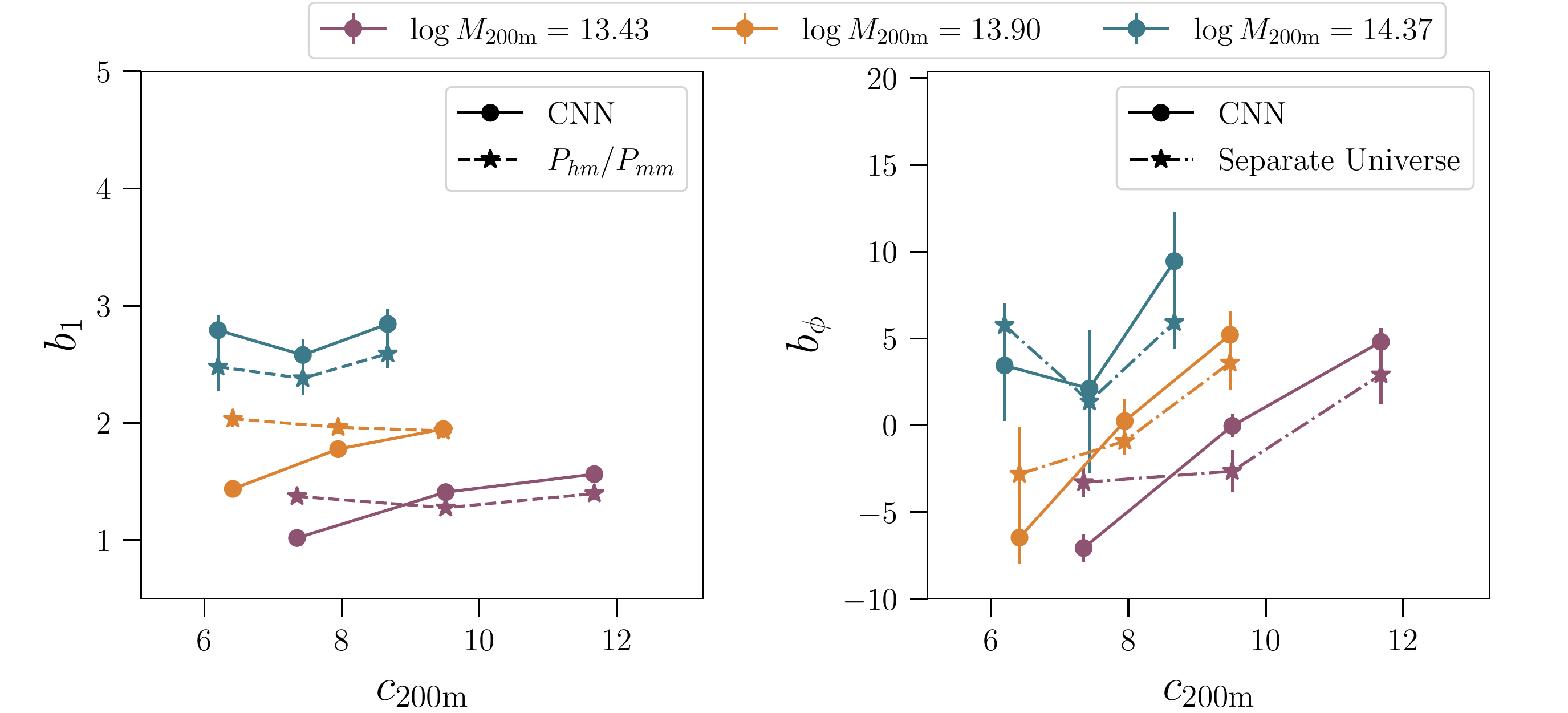}         
            \caption{Halo assembly bias in $b_1$ (\textit{left}) and $b_\phi$ (\textit{right}) as a function of concentration, derived from the $c-$CNN surrogate model for predicting halo concentration from the initial conditions. The three colours show the derived assembly bias for the three different mass bins; the three points in each colour show the measured assembly bias in each concentration bin for that halo mass bin. 
            For comparison with ``ground-truth'' values, we show also the $b_1$ obtained from the large-scale limit of $P_{hm}(k)/P_{mm}(k)$, and $b_\phi$ obtained with the separate-universe approach; these are shown as stars connected by dashed lines for $b_1$ and as stars connected by dot-dashed lines for $b_\phi$.}
    \label{fig:b1_assembly}
\end{figure*}

We now turn to halo assembly bias i.e. the fact that dark matter halos also exhibit bias in terms of secondary halo properties beyond mass. We use the halo concentration values predicted by the $c$-CNN model to derive the assembly bias in $b_1$ and $b_\phi$ as follows.

We follow a similar procedure to that described in Sec.~\ref{sec:halobias} for the bias as a function of mass alone.  We add constant density $\delta_{\rm L}$ and gravitational potential $f_\mathrm{NL}\phi_{\rm L}$ (or equivalently $\delta_{A_s}$) perturbations to each input density field. We consider the same perturbation values as in Sec.~\ref{sec:halobias}. This yields new scale radius predictions $r_\mathrm{s, \, pred}^{\delta}$, $r_\mathrm{s, \, pred}^{\phi}$, which we convert to halo concentration $c_\mathrm{pred}^{\delta}$, $c_\mathrm{pred}^{\phi}$; recall, these concentration predictions come from the $c$-CNN which also takes as input the mass predictions $M_\mathrm{pred}^{\delta}$, $M_\mathrm{pred}^{\phi}$ of the $M$-CNN with the injected perturbations. 
We measure the number density of halos as a function of predicted mass and concentration $n_h(M,c)$. We bin the halos by mass in three log-spaced bins between $\log_{10} M_\mathrm{min}=13.2$ and $\log_{10} M_\mathrm{max}=14.6$, and within each mass bin, we further split the halos by their concentration into three linearly spaced bins between the values of the 5th and 95th percentile of the concentration distribution in the test simulation (without any injected large-scale perturbations). Analogously to Eqs.~\eqref{eq:deltah} and \eqref{eq:deltahbphi}, we then estimate the change in the number density of halos in each mass and concentration bin. The bias parameters are measured using the same polynomial fitting procedure as in Sec.~\ref{sec:halobias}.

Figure \ref{fig:b1_assembly} shows the assembly bias in $b_1$ (left) and $b_\phi$ (right) as a function of concentration for the three halo mass bins (shown as three different colors). We show the mean and standard deviation of the distribution of assembly bias values obtained via bootstrap in each concentration bin. 
The results show modest assembly bias in $b_1$ as a function of concentration for our three mass bins. This is consistent with previous work \citep{Wechsler2006, Gao2007, Jing2007, 2017JCAP...03..059L, 2021JCAP...10..063L}, and with our direct measurements of $b_1$ (stars), for the same mass/concentration bins, from the large-scale limit of the ratio of the cross halo-matter power spectrum to the matter power spectrum (see Sec.~\ref{sec:sims}). 
The assembly bias derived by the CNN mildly deviates from the direct measurements, especially for halos with lowest concentration; these small differences are expected and indicative of the noise level present in our deep learning framework as the assembly bias signal is weak.
Nevertheless, the CNN is able to correctly recover the impact of the large-scale perturbations to the inputs on the final concentration, and to correctly derive the low level of halo assembly bias as a function of concentration. Since the assembly bias signal is already so low for $b_1$, we do not further investigate it for $b_2$ \citep{2021JCAP...10..063L}.

The right panel of Fig.~\ref{fig:b1_assembly} shows the same, but for $b_\phi$. The CNN predicts a strong halo assembly bias signal for all mass bins as a function of their concentration. This is in agreement with our direct $b_\phi$ measurements using separate-universe simulations (see Sec.~\ref{sec:sims}), as well as with the previous work of \cite{2023JCAP...01..023L} for similar mass halos. The values of $b_\phi$ increase with increasing value of concentration at fixed halo mass, and the magnitude of this effect is qualitatively similar for all halo masses considered in this work. For the two lowest mass bins, the values of $b_\phi$ increase from negative to positive with increasing concentration. That is, an increase in $A_{s}$ causes halos to form earlier and be more concentrated, which results in fewer low concentration objects, hence yielding $b_\phi < 0$. Conversely, the number of halos with high concentration increases if $A_{s}$ increases, hence the larger positive values of $b_\phi$. 

\subsection{The response of the concentration-mass relation}

\begin{figure}
     \centering
         \includegraphics[width=\columnwidth]{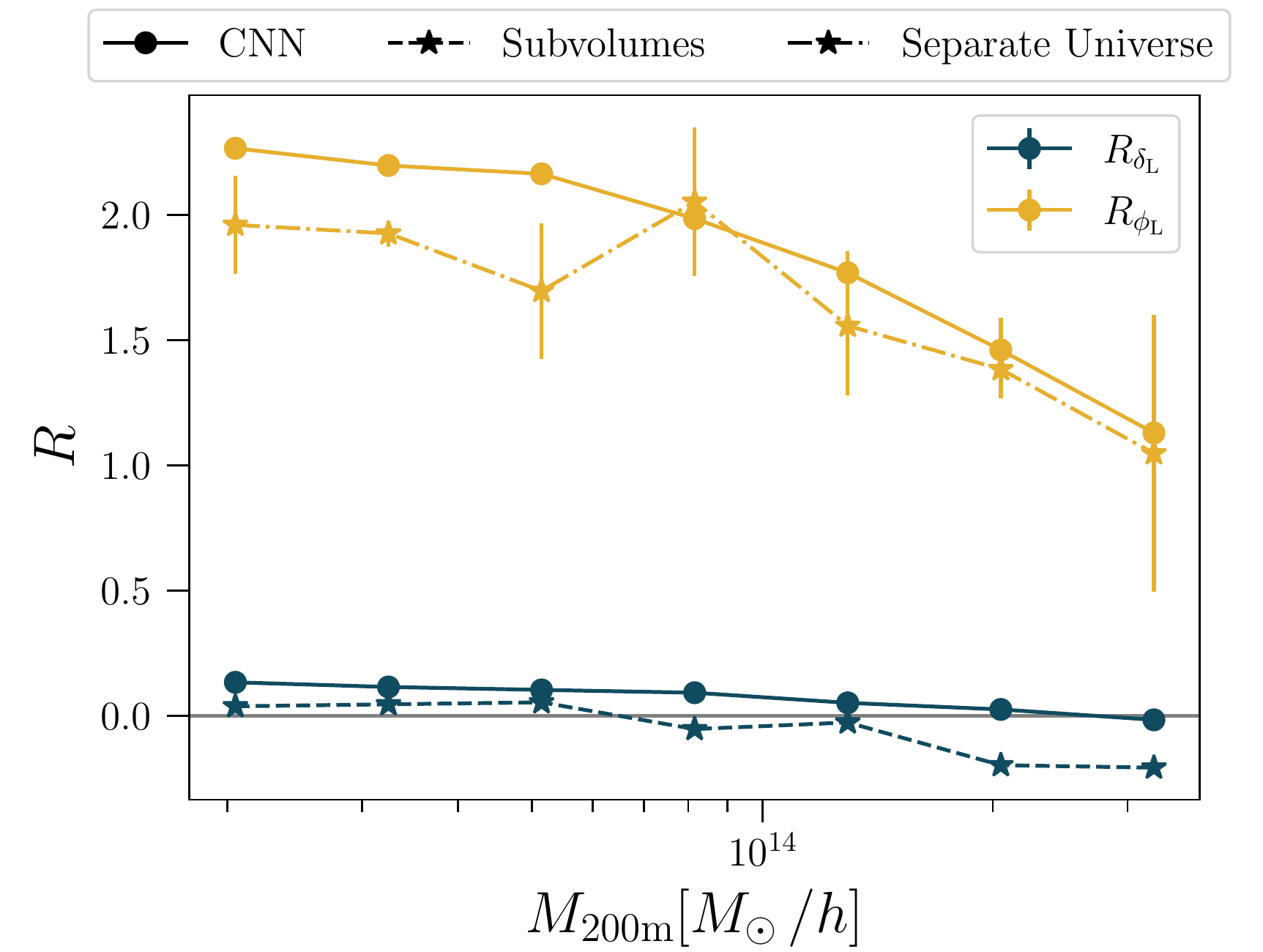}         
            \caption{The linear response of the mean concentration-mass relation to large-scale perturbations in the matter field, which we denote $R_{\delta_{\rm L}}$, and in the local PNG primordial gravitational potential, which we denote $R_{\phi_{\rm L}}$. The responses predicted by the CNN are comparable to those measured directly from simulations, using subvolumes of the test simulation for $R_{\delta_{\rm L}}$ and the separate-universe approach for $R_{\phi_{\rm L}}$.}
    \label{fig:cM_response}
\end{figure}
As a complementary viewpoint to our assembly bias results above for $b_1$ and $b_\phi$, we discuss now the response of the mean concentration-mass relation to a change in the background density $\delta_{\rm L}$ and in the gravitational potential $f_\mathrm{NL}\phi_{\rm L}$.

To do so, we measure the change in the mean concentration $\bar{c}$ over all halos in a given mass bin $M$ as
\begin{equation}
\delta{\bar{c}^X(M)} = \frac{\bar{c}^{X}(M)}{\bar{c}^0(M)} - 1,
\end{equation}
where $\bar{c}^{X}(M)$ is the mean over the predicted concentration values in the presence of the perturbation $X = \delta_{\rm L}$ or $X = \phi_{\rm L}$; $\bar{c}^{0}(M)$ is the same, but without any perturbations applied to the inputs. For every mass bin, we then fit the polynomial
\begin{equation}
\delta{\bar{c}^{X}(M)} = \sum_{i=1}^4 \frac{1}{n!}R_{n, X}(X)^n
\end{equation}
and infer the best-fit response parameters $R_{n, X}$ for the two types of large-scale perturbations, $X=\delta_{\rm L}$ and $X=\phi_{\rm L}$. 

Figure \ref{fig:cM_response} shows the linear responses $R_{\delta_{\rm L}} \equiv R_{1, \delta_{\rm L}}$ and $R_{\phi_{\rm L}} \equiv R_{1, \phi_{\rm L}}$ as a function of halo mass. As expected from the weak assembly bias in $b_1$ in Fig.~\ref{fig:b1_assembly}, the response of the concentration-mass relation to $\delta_{\rm L}$ is small i.e., at fixed mass, the mean halo concentration is not severely affected by large-scale matter perturbations. This in turn means that, at fixed mass, the concentration dependence of the halo abundance is not a strong function of $\delta_{\rm L}$, hence the weak dependence of $b_1(M, c)$ on $c$. Note that the mass and concentration of individual halos are functions of $\delta_{\rm L}$; the small values of $R_{\delta_{\rm L}}$ indicate however that the changes in mass are accompanied by changes in concentration that keep the relative concentration-mass values nearly unchanged. 
The blue stars in Fig.~\ref{fig:cM_response} show a direct measurement of $R_{\delta_{\rm L}}$ obtained by cross-correlating the mean concentration-mass relation in subvolumes of the test simulation with the mean density in the subvolumes.\footnote{Concretely, we split the test simulation into 64 subvolumes and compute (i) the mean matter density $\delta_{\rm L}$ and (ii) mean concentration-mass relation in each. We plot these 64 measurements against one another for each halo mass bin and measure $R_{\delta_{\rm L}}$ by fitting a line to them.} The two results agree in the recovery of a relatively weak response function, despite some small differences.

On the other hand, and also as expected from the strong assembly bias in $b_\phi$ in Fig.~\ref{fig:b1_assembly}, we see in Fig.~\ref{fig:cM_response} a strong response of the concentration-mass relation to perturbations in the local PNG primordial gravitation potential $\phi_{\rm L}$. Inside positive $f_\mathrm{NL}\phi_{\rm L}$ perturbations (i.e. in regions with enhanced variance of the primordial fluctuations, $\delta_{A_s}>0$), halos become in general both more massive and more concentrated; the large positive values of $R_{\phi_{\rm L}}$ indicate that the increase in concentration is more significant than the increase in mass. This in turn explains the strong halo assembly bias of $b_\phi$. For the halo mass range considered in this work, the response of the concentration-mass relation remains approximately the same for halos with mass $M \leq 10^{14} \mathrm{M}_\odot \, /h$, but starts to decline towards larger masses. We also compare our results for $R_{\phi_{\rm L}}$ to direct measurements obtained using separate-universe simulations (yellow stars).\footnote{These are obtained by finite-differencing the concentration-mass relation w.r.t.~$\delta A_s$ using our complementary set of separate-universe simulations (see Sec.~\ref{sec:sims}). This is analogous to the measurements of $b_\phi$ obtained by finite-differencing the halo abundance.} We find that the decline in $R_{\phi_{\rm L}}$ for halos with $M>10^{14} M_\odot \, /h$ is in perfect agreement between the two methods, but the amplitude $R_{\phi_{\rm L}}$ on smaller mass scales is slightly over-estimated by the CNN.

\section{Conclusions}
\label{sec:conclusions}
We have presented a deep learning framework that links halo bias to the properties of the initial density field that determine the final mass and concentration of dark matter halos. Our goal was to show that the same Lagrangian properties that are relevant for halo formation also determine halo bias when coupled to changes in the initial large-scale environment. We make use of the exact definition of halo bias as the response of the abundance of halos to large-scale perturbations. Our framework consists of a deep learning model that maps the initial density field around each protohalo centre to the final mass and concentration of the resulting $z=0$ halo. Once trained, the model is then used as a surrogate to derive the halo bias parameters using the response of the predicted mass and concentration to large-scale perturbations injected in the initial conditions (Fig.~\ref{fig:illustration}). It should be emphasized that no explicit knowledge about halo bias was provided to the model during training. 

We focused on the linear and quadratic density bias parameters, $b_1$ and $b_2$, which measure the response of halo formation to large-scale perturbations in the matter density field $\delta_{\rm L}$, and on the linear PNG bias parameter $b_\phi$, which measures the response to perturbations in the primordial gravitational potential induced by local PNG $f_{\rm NL}\phi_{\rm L}$. Our framework can be straightforwardly extended to other large-scale perturbations including for example perturbations to the large-scale tidal field to measure tidal bias parameters. 

We find that the halo bias as a function of halo mass predicted by our CNN framework is in good agreement with that measured directly from $N$-body simulations (Fig.~\ref{fig:b1}). Our results match fitting formulae for $b_1(M)$ and $b_2(b_1)$ calibrated to $N$-body simulations, and the expected amplitude and linear shape of the $b_\phi(b_1)$ relation.

We then used our framework to predict the assembly bias of halos in terms of their concentration (Fig.~\ref{fig:b1_assembly}). We find that, at fixed mass, and in accordance with the expectation from previous works for similar halo masses, halos of different concentration exhibit weak assembly bias in $b_1$, but a strong assembly bias in $b_\phi$. In local PNG cosmologies, inside gravitational potential perturbations the abundance of halos with high concentration is strongly enhanced, whereas the number of halos with low concentration is suppressed. We also investigated the responses of the concentration-mass relation (Fig.~\ref{fig:cM_response}), which provide another viewpoint on the assembly bias results: the mean concentration of halos of a given mass is strongly enhanced by positive $f_\mathrm{NL}\phi_{\rm L}$ perturbations, but is left approximately unchanged by $\delta_{\rm L}$ perturbations. This result can be explained by the differential impact that these two types of perturbations have on the mass and concentration of individual halos. Our CNN results for assembly bias also match well the result obtained directly from simulation measurements. Our deep learning approach shows overall that the features extracted by the CNN to predict halo concentration also correctly determine assembly bias, i.e. how secondary properties of halos (such as concentration) respond to large-scale perturbations at fixed halo mass.

Our framework is closely related to analytic implementations of the PBS formalism. Local features of the initial density field are extracted from the initial conditions and used to infer final halo properties; in our framework this is done by the CNN, whereas analytic approximations typically take spherical overdensities to be the relevant features.\footnote{Note however that, compared to analytic approximations, we additionally provide information about the centre-of-mass of the protohalo region.} The presence of a large-scale perturbation modulates those initial conditions features in a way that affects the halo formation process: the response to large-scale perturbations predicted by the CNN is in agreement with that measured directly from simulations. This implies that the CNN has learnt features of the initial conditions that provide a good description of halo formation, thus providing accurate predictions for halo statistics that go beyond what the model was directly trained on. Our work opens a new window for discovering the features of the protohalo region that are responsible for the bias of halos, including their assembly bias signal.

Going forward, we must interpret the features learnt by the CNN and explain them in terms of physical aspects of the initial density field. In future work, we will adopt an interpretability technique developed in previous work \citep{Lucie-Smith2020}, where certain aspects are removed from the inputs to test the impact of this on the accuracy of the final predictions. We will quantify the effect of modifying properties of the initial density field — e.g. removing its anisotropic component or modifying the matter distribution within the protohalo region — on the accuracy of the assembly bias predictions. This will also allow us to test the validity of other proposed analytic or numerical solutions, where for example halo bias is thought to be related to the curvature of the initial peaks \citep{Dalal2008} or the local tidal environment \citep{Paranjape2018, Ramakrishnan2019}. Our framework has the potential of shedding light on the origin of halo assembly bias, provided that the features learnt by the CNN can be interpreted.

\section*{Acknowledgements}
LLS thanks Raul Angulo, Corentin Cadiou and Andrew Pontzen for useful discussions. LLS, AB and FS acknowledge the hospitality of the Aspen Center for Physics, which is supported by National Science Foundation grant PHY-1607611. The participation of LLS at the Aspen Center for Physics was supported by the Simons Foundation. AB acknowledges support from the Excellence Cluster ORIGINS which is funded by the Deutsche Forschungsgemeinschaft (DFG, German Research Foundation) under Germany's Excellence Strategy - EXC-2094-390783311.

\section*{Data Availability}
The data underlying this article will be shared upon reasonable request to the corresponding author.



\bibliographystyle{mnras}
\bibliography{biaspaper} 






\bsp	
\label{lastpage}
\end{document}